# Spatial–Molecular Information Analysis Based on Histopathological Geometry: KL-Divergence Decomposition of Cell Distribution and Location-Dependent Expression States in PD-1 Immunohistochemistry

Tatsuaki Tsuruyama, M.D., Ph.D.
Department of Drug Discovery Medicine, Graduate School of Medicine,
Kyoto University, Kyoto, Japan
Department of Physics, Graduate School of Medicine,
Tohoku University, Sendai, Japan

Correspondence: (email:tsuruyam@kuhp.kyoto-u.ac.jp)

## Abstract

Spatial biology enables molecular expression to be analyzed together with the location of cells in tissue. However, it is often difficult to distinguish whether cells preferentially accumulate near a histopathological boundary from whether their molecular expression states vary with that location. To establish informational scientific framework, we define boundary distance as $R$ and a binary molecular expression state as $G$, and compare the observed joint distribution $P(R, G)$ with $Q(R)P(G)$, where $Q(R)$ is a null distribution determined by tissue geometry. The resulting Kullback-Leibler divergence decomposes into $D_R$, which quantifies deviation of the spatial distribution of cells from the geometric null, and $I(R; G)$, which measures how strongly molecular expression state depends on boundary distance. Idealized simulations and semi-synthetic validation using geometry reconstructed from a PD-1 IHC image confirmed that these two effects can be generated separately and recovered by the corresponding terms. Plug-in estimators showed positive finite-sample bias, supporting permutation-based inference. We then applied the framework to five publicly available PD-1 lung cancer IHC images from the Human Protein Atlas. In pooled analysis, $I(R; G) = 0.0302$ nats and was significant in a stratified within-image label-permutation test ($p = 0.00040$). This framework provides a quantitative way to distinguish where cells are located from how their molecular expression states vary with tissue location. Larger cohorts, independent datasets, additional cancer types, and other molecular markers will be required to establish generalizability and clinical utility.

## 1. Introduction

Spatial transcriptomics and multiplex tissue imaging provide a basis for linking molecular expression to tissue coordinates and for analyzing heterogeneity in the tumor microenvironment [3]. Recent studies have extended spatial analysis beyond cell-type density and proximity to include entropy- and mutual-information-based analyses across distance scales [4], analyses that explicitly use the tumor boundary as a spatial axis [5], and approaches that evaluate cell-tumor spatial relationships under null hypotheses that account for tissue geometry [6]. These studies demonstrate that spatial organization within tumors carries biological information; however, the effect of cellular placement near a boundary and the effect of preferential molecular expression states at a given distance can be conflated depending on the analytical objective.

Here, we distinguish these two effects within a single information-theoretic framework. We define $R$ as the distance from a histopathologically defined boundary and $G$ as a coarse-grained molecular expression state, and compare the observed joint distribution $P(R, G)$ with $Q(R)P(G)$, where $Q(R)$ is a distance null distribution determined solely by tissue-image geometry and $P(G)$ is the overall state distribution of the specimen. This KLD decomposes exactly into the placement term $D_R$ and the state-coupling term $I(R; G)$ [1,2]. The novelty of the present study lies not in KLD or mutual information themselves, but in incorporating a pathologically defined boundary and tissue geometry into the null model and evaluating placement and expression-state effects within the same decomposition.

We first assessed the separability and finite-sample behavior of $D_R$ and $I(R; G)$ using an idealized simulation with a circular cancer nest and a semi-synthetic validation using cancer-nest geometry reconstructed from an actual PD-1 IHC image [7-9]. We then applied the framework to publicly available PDCD1 (PD-1) lung carcinoma IHC images from the Human Protein Atlas (HPA) [10,11] to evaluate the statistical coupling between cancer-nest boundary distance and PD-1-positive/negative state. Finally, to relate specimen-level $I(R; G)$ to local distance dependence, we visualized the distance gradient of the local KLD on the histopathology image.

## 2. Coarse-Grained Variables and Observational State Space

### 2.1 Microscopic State and Coarse-Graining Map

Let the microscopic state of the tissue be denoted by ω. The state ω may include cell position, cell type, gene expression, protein expression, and the states of neighboring cells. We consider an observation map z that transforms this high-dimensional state into coarse-grained variables: $z: \omega \to (R, G)$. Here, $R$ is defined as the shortest distance from a cell position to a histopathological boundary. $G$ denotes a coarse-grained molecular expression state (gene or protein); for example, for a binary expression state $G \in \{0,1\}$, $G = 1$ denotes positive expression and $G = 0$ denotes negative expression. The resulting coarse-grained probability distribution is denoted by $P(R, G)$. The marginal distributions are given by:

$$P(R) = \sum_{g} P\,(R, g), \qquad P(G = g) = \sum_{r} P\,(r, g)$$

## 3. Geometry-Conditioned Decomposition of Spatial-Molecular Expression Information

### 3.1 Definition of the Geometry-Derived Null Distribution *Q*(*R*)

Let Ω denote the analyzable region and $\partial C$ a histopathologically defined boundary. For each analyzable position $x \in \Omega$, the shortest distance to the boundary is $R = d(x, \partial C)$. The distance is discretized into bins r. If $A_r$ denotes the analyzable area (number of pixels) contained in bin r, the geometry-derived null distribution is defined as $Q(r) = \frac{A_r}{\sum_r A_r}$. Because $Q(R)$ does not use the molecular expression state $G$, it is determined solely by tissue geometry. Thus, $Q(R)$ represents the boundary-distance distribution expected if cells were placed uniformly at random within the analyzable region.

### 3.2 Total Spatial-Expression Divergence and Its Decomposition

Let $P(R, G)$ denote the empirical joint distribution of observed cells or spots. If n(r,g) is the number of observations in distance bin r and molecular expression state g, and N is the total number of observations, then

$$P(r, g) = \frac{n(r, g)}{N}$$

and the marginal distributions are $P(r) = \sum_g P\,(r, g)$ and $P(g) = \sum_r P\,(r, g)$. Here, $P(R)$ is the observed boundary-distance distribution of all cells or spots combined, irrespective of expression state. Using these distributions, we define the total Kullback-Leibler divergence as follows.

$$D_{RG} = D_{\mathrm{KL}}\big(P(R, G) \parallel Q(R)P(G)\big)$$

Equivalently,

$$D_{RG} = \sum_{r,g} P\,(r, g)\log\frac{P(r, g)}{Q(r)P(g)}$$

This quantity measures how far the observed spatial-expression joint distribution $P(R, G)$ deviates from $Q(R)P(G)$, the product of the spatial null distribution and the overall expression-state distribution. Accordingly, $D_{RG}$ contains both the spatial bias of cellular placement and the distance-dependent change in the composition of expression state *G*. The logarithmic term can be decomposed as follows.

$$\log\frac{P(r, g)}{Q(r)P(g)} = \log\frac{P(r)}{Q(r)} + \log\frac{P(r, g)}{P(r)P(g)}$$

Averaging both sides with respect to $P(r, g)$ yields the following identity.

$$D_{RG} = D_R + I(R; G)$$

where

$$D_R = D_{\mathrm{KL}}\big(P(R) \parallel Q(R)\big) = \sum_r P\,(r)\log\frac{P(r)}{Q(r)}$$

represents the geometric bias of cellular placement without distinguishing positive from negative expression states. In other words, it quantifies how far the boundary-distance distribution $P(R)$ of all cells deviates from the area-corrected random-placement reference $Q(R)$ expected from image geometry.

By contrast,

$$I(R; G) = \sum_{r,g} P\,(r, g)\log\frac{P(r, g)}{P(r)P(g)}$$

is the mutual information between boundary distance *R* and the coarse-grained molecular expression state *G*. It quantifies either how much knowing *R* reduces uncertainty in *G* or, equivalently, how strongly the expression-state composition in each distance bin differs from the overall composition. The mutual information can also be written as a distance-weighted average of local KLDs.

$$I(R; G) = \sum_r P\,(r)\, D_{\mathrm{KL}}\big(P(G \mid R = r) \parallel P(G)\big)$$

Thus, $D_R$ represents the placement effect—where cells are located—whereas $I(R; G)$ represents the state effect—which expression state cells tend to adopt at a given distance. For IHC, $D_R$ is the deviation of the boundary-distance distribution of all annotated cells without regard to positive/negative status, whereas $I(R; G)$ measures how the positive/negative composition changes with boundary distance.

For a binary positive/negative state $G$, let $p_r = P(G = 1 \mid R = r)$ denote the positive fraction in distance bin r and $p = P(G = 1)$ the overall positive fraction. The mutual information can then be written as

$$I(R;G) = \sum_r P(r)\left[p_r \log\frac{p_r}{p} + (1-p_r)\log\frac{1-p_r}{1-p}\right]$$

The first term represents the contribution of enrichment or depletion of the positive state in distance bin r, and the second term represents the contribution of the negative state. Therefore, $I(R;G)$ is not simply the number or density of positive cells or the distance distribution of positive cells alone; rather, it measures how the positive/negative composition itself varies along $R$.
This decomposition distinguishes placement and state effects. If $D_R \approx 0$ and $I(R;G) > 0$, the overall cell-distance distribution is close to the geometric null distribution while the expression-state composition remains distance dependent. Conversely, if $D_R > 0$ and $I(R;G) \approx 0$, cellular placement is non-random relative to the boundary but the positive/negative state composition is not distance dependent. If both are positive, placement bias and state coupling coexist.

**3.3 Simulation Validation in an Idealized Tissue Geometry**
To examine the behavior of $D_R$ and $I(R;G)$ under known ground-truth conditions, we performed an idealized simulation in a square analysis field with a circular cancer nest at the center (Figure 1A). $Q(R)$ was constructed from the tissue geometry. Boundary-dependent cellular placement was controlled by parameter α, and boundary-distance-dependent expression-state composition was controlled by parameter β. The cellular placement distribution and binary expression state were generated as follows.

$$P_\alpha(r) = \frac{Q(r)e^{-\alpha r_{\text{scaled}}}}{\sum_{r'} Q(r')e^{-\alpha r'_{\text{scaled}}}}$$
$$P(G = 1 \mid R = r) = \text{logit}^{-1}[c + \beta(0.5 - r_{\text{scaled}})]$$

For each condition, c was adjusted so that the overall positive fraction matched that of the empirical pooled data (102/567). Generating known spatial structures in simulation to evaluate analytical methods is widely used in spatial-omics method development and power analysis [7-9].
Figure 1A shows four conditions: Null ($\alpha = 0, \beta = 0$), Placement only ($\alpha = 2, \beta = 0$), State only ($\alpha = 0, \beta = 3.5$), and Both ($\alpha = 2, \beta = 3.5$). At the population level, the ground-truth values were $D_R = 0$ and $I(R;G) = 0$ for Null; $D_R = 0.0905$ and $I(R;G) = 0$ for Placement only; $D_R = 0$ and $I(R;G) = 0.0372$ for State only; and $D_R = 0.0905$ and $I(R;G) = 0.0296$ for Both (Figure 1B). Thus, $D_R$ became selectively positive when only cellular placement was altered, whereas $I(R;G)$ became selectively positive when only state composition was made distance dependent.
When α and β were varied continuously (Figure 1C), $D_R$ increased with α and was independent of β. By contrast, $I(R;G)$ increased primarily with β. However, $I(R;G)$ is a distance-weighted average of the local KLD, and contributions from each distance are weighted by $P(R)$. Therefore, even at fixed β, $I(R;G)$ can change when α alters $P(R)$. The simulation therefore demonstrates separability—not statistical independence—of the two components: placement and state effects can be generated as distinct ground truths and recovered by their corresponding terms.

$N = 567I(R;G)$Finite-sample behavior is shown in Figure 1D. Even under the Null condition, the plug-in estimator obtained by directly substituting empirical frequencies into the information formulas took small positive values, and this bias decreased as the number of cells increased (Figure 1D, upper panel). This finite-sample bias is a known problem in mutual-information estimation [12,13]. In the label-permutation test, cell positions and boundary distances R were held fixed while only the G labels were shuffled; this procedure estimates how large I(R;G) can arise by chance under the same sample size and distance structure if R and G are unrelated. At N=567, the rejection rates were 0.044 for Null, 0.992 for State only, and 0.996 for Both (Figure 1D, lower panel). These results support evaluating significance against a permutation null constructed under the same sample size and distance structure rather than judging significance from the absolute value of I(R;G) alone.

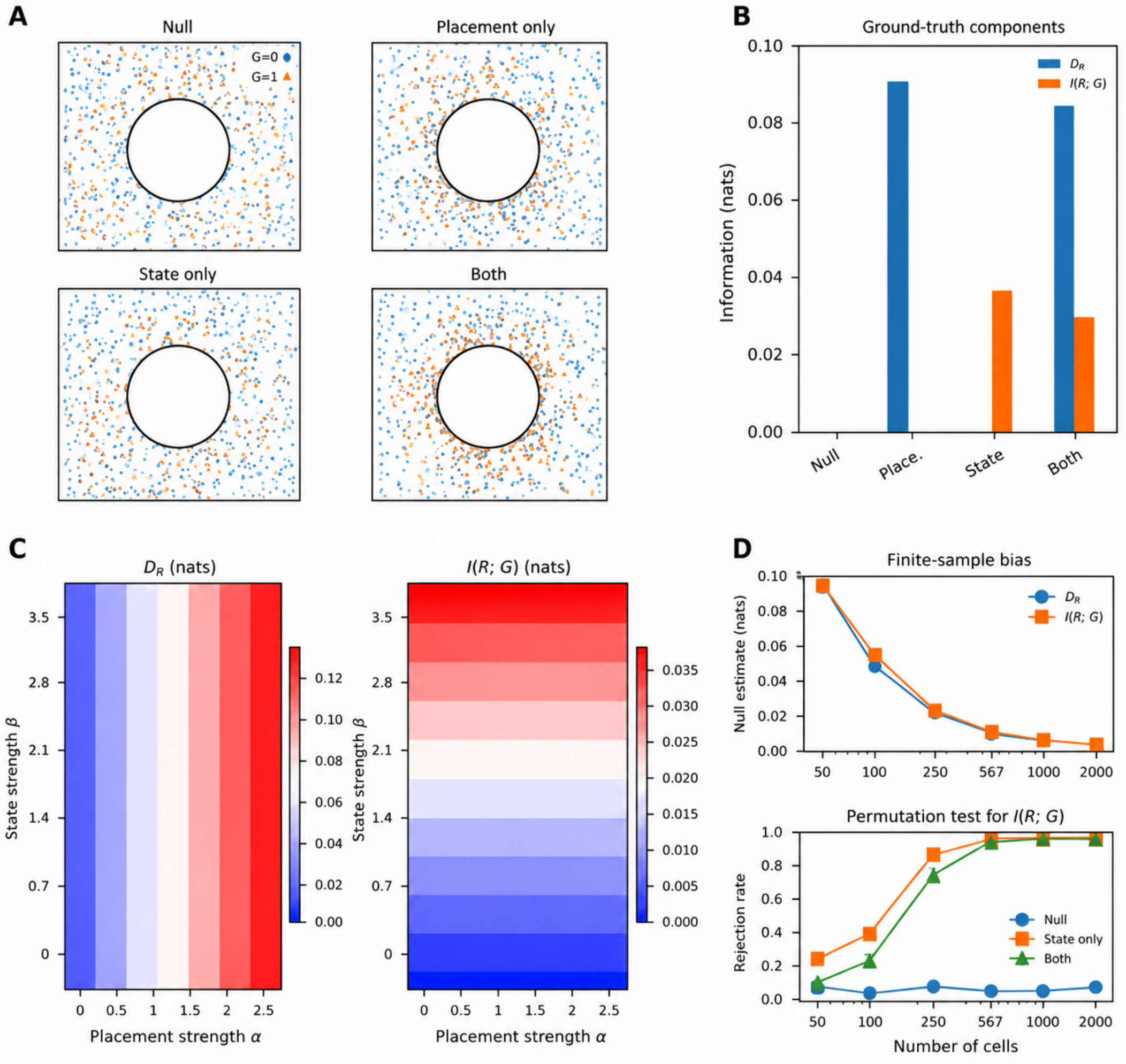

**Figure 1.** Simulation validation of $D_R$ and $I(R;G)$ in an idealized tissue geometry. (A) Four conditions with a circular cancer nest at the center. Blue circles denote $G = 0$ and orange triangles denote $G = 1$. Null has no distance dependence in either placement or state; Placement only introduces distance dependence in placement alone; State only introduces distance dependence in expression-state composition alone; Both introduces both effects. (B) Population-level ground-truth values for each condition. Only $D_R$ is positive in Placement only, whereas only $I(R;G)$ is positive in State only. (C) Heatmaps obtained by varying placement-bias strength α and state-coupling strength

β. Colors range from blue (low) to red (high). (D) Finite-sample behavior. The upper panel shows plug-in estimates under the Null condition; the lower panel shows rejection rates for the label-permutation test of $I(R;G)$. Information is reported in nats using natural logarithms. Parameter values were chosen to illustrate the decomposition behavior and were not fitted to the empirical data.

## 4. Application to PD-1 IHC Images

### 4.1 Study Design and Analytical Workflow

After confirming the basic behavior of the decomposition in the idealized simulation, we performed a semi-synthetic validation using fixed cancer-nest geometry reconstructed from an annotated PD-1 IHC image to assess robustness to realistic tissue shape. We then applied the framework to five publicly available HPA PDCD1 (PD-1) lung carcinoma IHC images and evaluated the coupling between cancer-nest boundary distance $R$ and PD-1 positive/negative state $G$ as $I(R;G)$ [10,11]. Finally, using a representative field, we visualized the local state information $J(R)$ and its distance gradient to relate specimen-level $I(R;G)$ to local patterns on the histopathology image.

### 4.2 Semi-Synthetic Validation Using Real Tissue Geometry

Tissue geometry was reconstructed from the cancer-nest boundaries drawn on an annotated PD-1 IHC image, and $Q(R)$ was determined from the distribution of distances between analyzable pixels and the reconstructed cancer nests (Figure 2A,B). With this geometry fixed, placement bias α and expression-state coupling β were manipulated independently to generate four conditions: Null, Placement bias only, State coupling only, and Placement + state coupling (Figure 2C). The overall positive fraction was fixed at 102/567, matching the empirical pooled data. α and β were illustrative values chosen to make the decomposition behavior clear and were not fitted to the empirical data.

At the population level, the ground-truth values were $D_R = 0.09536$ and $I(R;G) = 0$ for Placement bias only; $D_R = 0$ and $I(R;G) = 0.03869$ for State coupling only; and $D_R = 0.09536$ and $I(R;G) = 0.02814$ for Placement + state coupling. Thus, even when $Q(R)$ reflected an irregular real-tissue geometry, placement bias and changes in expression-state composition independent of placement density could be generated and recovered as distinct components (Figure 2D).

When $N = 567$ cells were sampled and the procedure was repeated 2,000 times, the mean plug-in estimates under the Null condition were $D_R = 0.00801$ and $I(R;G) = 0.00825$ nats, reproducing a small positive finite-sample bias. By contrast, $D_R$ increased clearly under Placement bias only and $I(R;G)$ increased clearly under State coupling only, in accordance with the corresponding ground truths. This semi-synthetic validation shows that the decomposition behavior is retained not only in idealized circular geometry but also in an irregular geometry resembling real tissue.

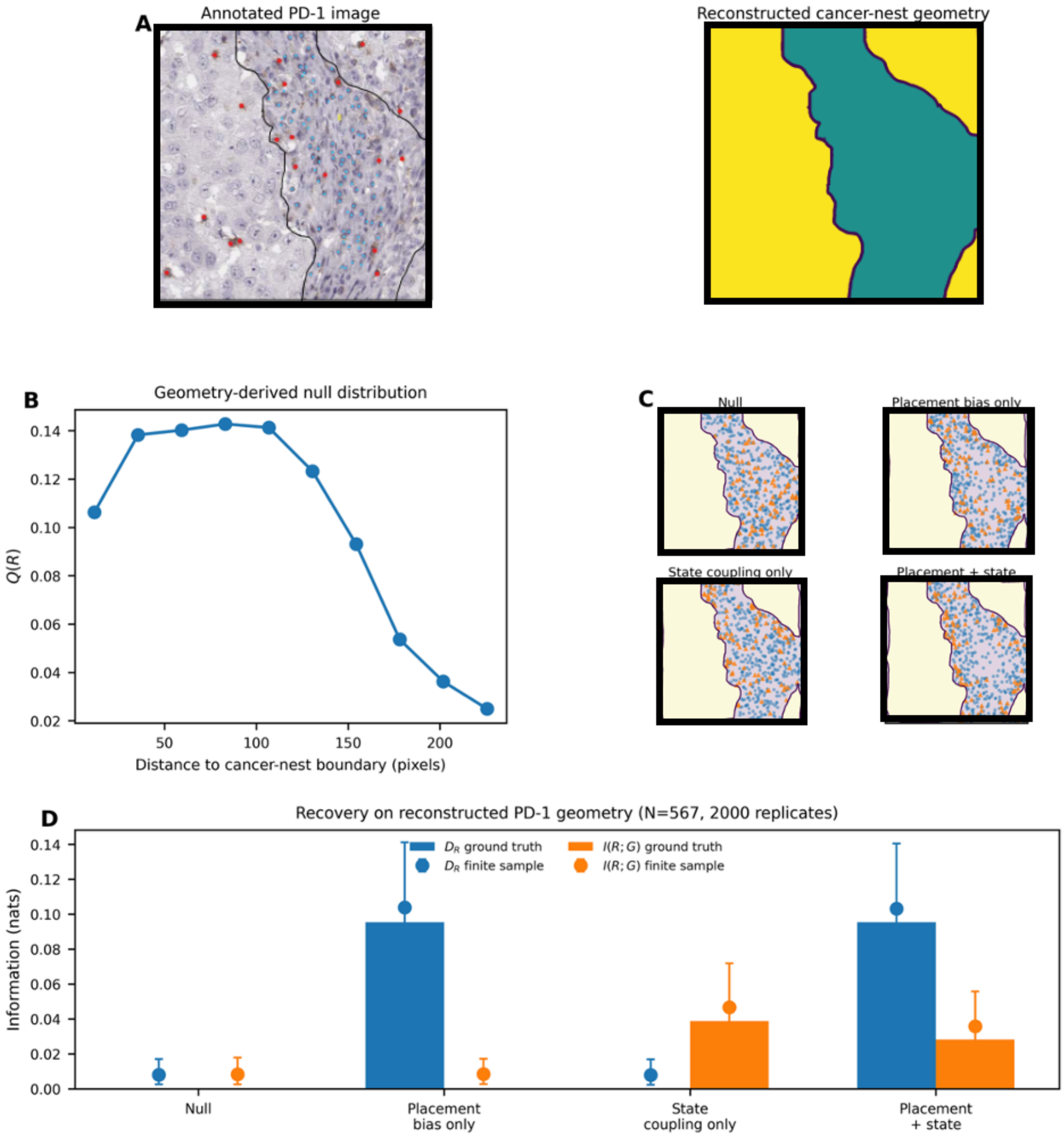

**Figure 2.** Semi-synthetic validation using cancer-nest geometry reconstructed from an annotated PD-1 IHC image. (A) Left, the annotated PD-1 IHC image used for analysis; right, cancer-nest regions reconstructed from the manually drawn boundaries. The central non-tumor region was used as the analyzable region (Lung (T-28000), Adenocarcinoma, NOS (M-81403) Patient id: 1907, , https://www.proteinatlas.org/ENSG00000188389-PDCD1/cancer/lung+cancer#img). (B) Geometry-derived null distribution $Q(R)$, calculated from the fraction of analyzable pixels in each distance bin from the reconstructed cancer-nest boundary. (C) Representative semi-synthetic examples for Null, placement bias only, expression-state coupling only, and placement + expression-state coupling. (D) Finite-sample recovery of $D_R$ and $I(R;G)$. Bars show population ground-truth values, points show mean plug-in estimates across 2,000 replicates, and error bars indicate the 2.5th-97.5th percentile

range. The parameters were chosen for illustration and were not fitted to the empirical PD-1 data. Original image in panel A: Human Protein Atlas (Image credit: Human Protein Atlas) [10,11].

### 4.3 Boundary-Distance Information Analysis in Five Independent Cases

We next applied the framework to five independent cases selected from publicly available HPA PDCD1 (PD-1) lung carcinoma IHC images, using one representative field per case [10,11]. Selection criteria were the presence of identifiable PD-1-positive cells, cancer nests occupying approximately ≥40% of the image area, identifiable intratumoral or peritumoral stroma, and an analysis field not excessively dependent on the edge of the registered image. Cancer-nest boundaries and PD-1-positive and -negative cells were annotated in each field, and $D_R$, $I(R;G)$, and $D_{RG}$ were calculated (Table 1). Because HPA requires citation of the atlas itself, direct citation of the specific image/gene/data used, and an image credit when images are displayed, the HPA patient/image ID, antibody ID, and direct URL for each case should be provided in a supplementary table at preprint submission [11].

In the pooled analysis of the five cases, 102 PD-1-positive and 465 PD-1-negative cells were identified. The median boundary distance was 52.6 pixels for positive cells and 99.3 pixels for negative cells, and positive cells were located significantly closer to the boundary (one-sided Mann-Whitney U test, $p = 1.93 \times 10^{-9}$). The pooled mutual information was $I(R;G) = 0.0302$ nats and was significant in a stratified permutation test in which PD-1 labels were shuffled within each image ($p = 0.00040$). The geometric bias of the distance distribution of all annotated cells was $D_R = 0.0965$, and the total divergence was $D_{RG} = D_R + I(R;G) = 0.1267$. Although permutation p values differed among cases, the median boundary distance of PD-1-positive cells was shorter than that of negative cells in all five cases (Table 1). Thus, in this dataset, a boundary-distance-dependent change in the PD-1 positive/negative composition was detected in addition to the overall placement bias of cells.

### 4.4 Visualization of Local State-Information Gradients

Because $I(R;G)$ summarizes the specimen-level statistical coupling as a single value, we calculated local state information $J(R)$ to identify the distance ranges at which this coupling changes (Figure 3A). The local positive fraction $q(R) \approx P(G = 1 \mid R)$ was estimated by Gaussian-kernel smoothing, and the binary KLD relative to the overall positive fraction $p = P(G = 1)$ was defined as follows.

$$J(R) = q(R)\log\frac{q(R)}{p} + [1 - q(R)]\log\frac{1 - q(R)}{1 - p}$$

The curve in Figure 3A is not the mutual information itself, but the distance-dependent local KLD that contributes to $I(R;G)$.

Figure 3B shows histograms of boundary distance for PD-1-positive and -negative cells in the representative field, whereas Figure 3C compares the observed $I(R;G)$ with its permutation null distribution for the same field. This representative field contained 19 positive and 95 negative cells, with an observed $I(R;G) = 0.0843$ nats and permutation $p = 0.0142$. The ECDF in Figure 3D shows, over the full distribution, that positive cells are shifted toward shorter distances. Figure 3E shows that the positive fraction is higher near the boundary and decreases at greater distances. Together, these panels illustrate that $I(R;G)$ captures a distance-dependent change in positive/negative composition rather than simply the number of positive cells. Figure 3 was generated from a representative field analyzed with the inside-zero setting for local visualization and is not the numerical source of the five-case summary in Table 1.

In Figure 3F, the original PD-1 IHC image is overlaid with cancer-nest boundaries, PD-1-positive and -negative markers, and white arrows based on the local information gradient. Black lines indicate cancer-nest boundaries, red points indicate $G = 1$ (PD-1 positive), and blue points indicate $G = 0$ (PD-1 negative). For each positive cell outside a cancer nest, the direction toward the nearest boundary point was defined geometrically, and arrow length and thickness were rescaled according to the magnitude of the local information gradient shown below.

$$F_{\text{info}}(R) = \left| \frac{dJ(R)}{dR} \right|$$

Accordingly, the white arrows visualize a boundary-directed information gradient that highlights locations where the local positive/negative composition changes sharply with boundary distance. They do not represent the actual direction or velocity of cell migration, a physical force, or an energy flux. Using Figure 3A-F together relates the global statistic $I(R;G)$ to complementary local representations: local KLD, distance distributions, permutation-based significance, ECDFs, distance-dependent positive fractions, and boundary-directed information gradients on the original histopathology image.

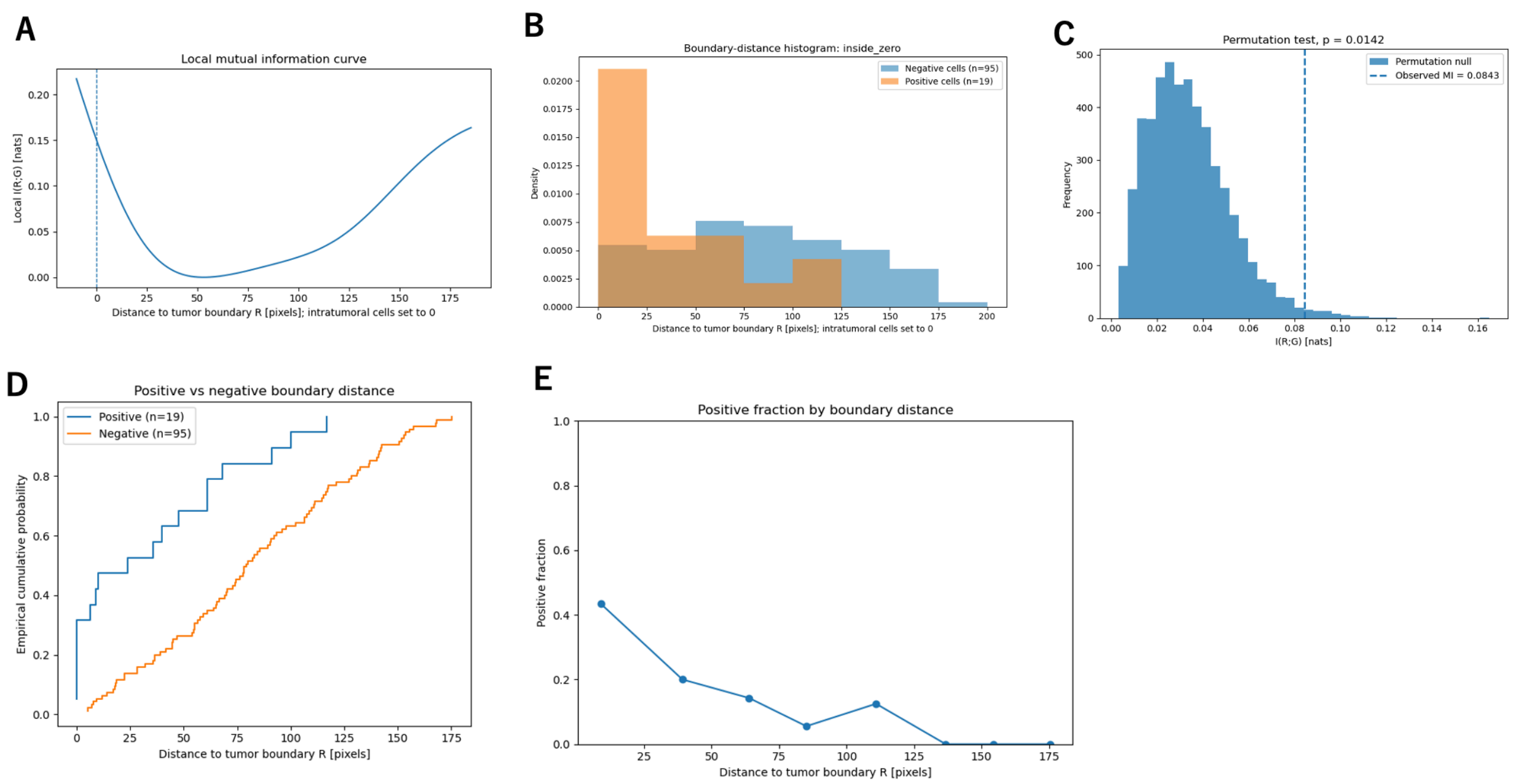

F

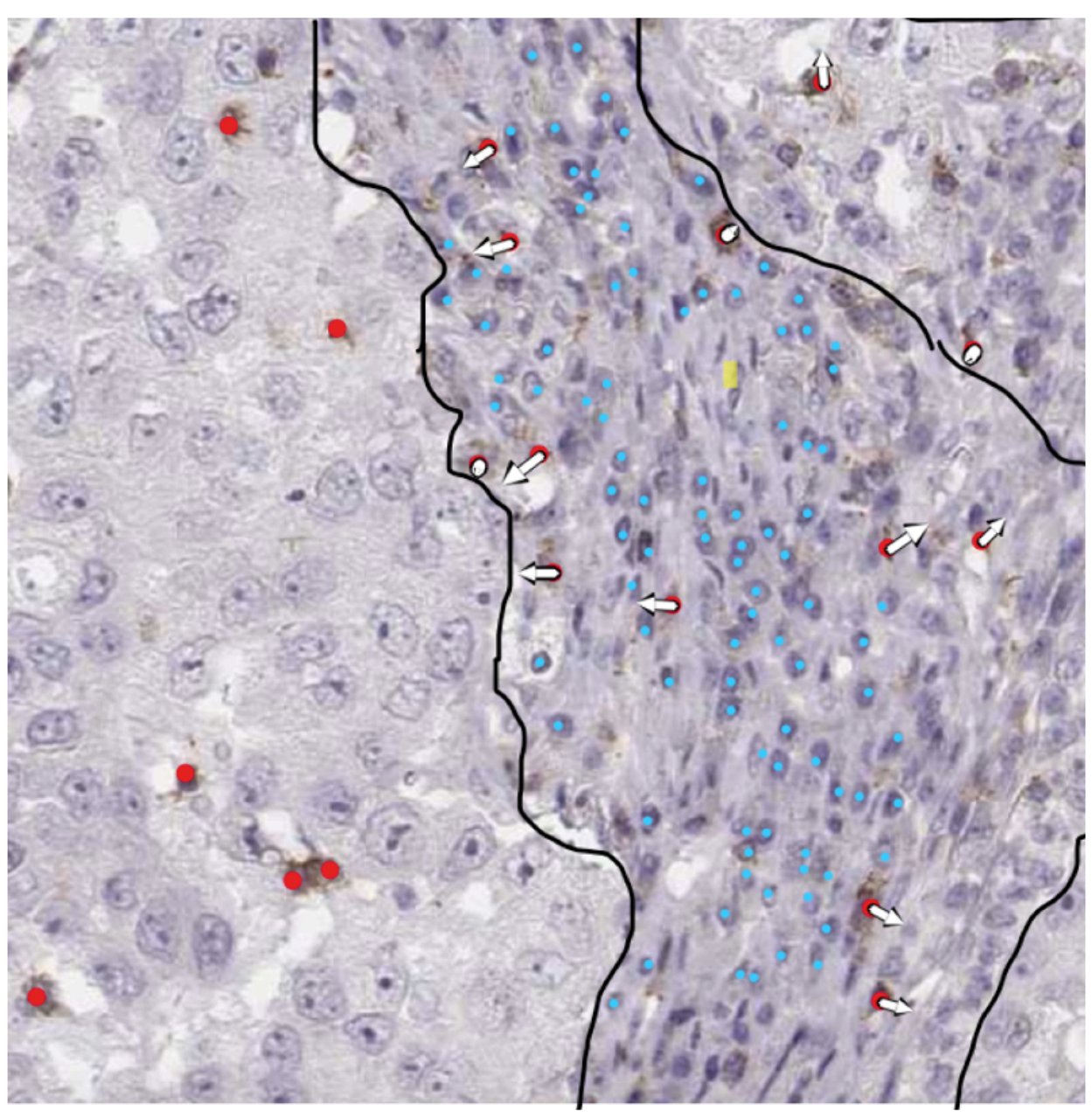

**Figure 3.** Visualization of cancer-nest boundary-distance dependence and local state information in a representative PD-1 IHC field. (A) Local state information $J(R)$ along distance $R$ from the cancer-nest boundary. The local positive fraction $q(R)$ was estimated by Gaussian-kernel smoothing and converted to a binary KLD relative to the overall positive fraction p. The vertical dashed line denotes $R = 0$. Under the inside-zero visualization setting, cells inside cancer nests were assigned $R = 0$. The label "Local mutual information curve" in the panel mathematically corresponds to the local KLD $J(R)$.
(B) Boundary-distance histograms for PD-1-positive and -negative cells. The representative field contained 19 positive and 95 negative cells; positive cells were shifted toward shorter distances.
(C) Label-permutation test for $I(R; G)$ between boundary distance $R$ and PD-1 state $G$. Cell positions and boundary distances were held fixed while only $G$ labels were permuted. The observed value was $I(R; G) = 0.0843$ nats with $p = 0.0142$. The dashed line indicates the observed value and the histogram the null distribution.
(D) ECDFs of cancer-nest boundary distance for PD-1-positive and -negative cells. The positive-cell curve lying toward the upper left indicates that positive cells are distributed at shorter distances.
(E) PD-1-positive fraction by boundary-distance bin. The fraction is higher near the boundary and tends to decline at greater distances.
(F) Original PD-1 IHC image overlaid with cancer-nest boundaries (black), PD-1-positive cells (red, $G = 1$), PD-1-negative cells (blue, $G = 0$), and white arrows. Arrow direction is the geometrically defined direction from each positive cell outside a cancer nest toward the nearest cancer-nest boundary, whereas arrow length and thickness are based on $F_{\text{info}}(R) = \left|\frac{dJ(R)}{dR}\right|$. This is a visualization of a boundary-directed information gradient and does not represent actual force or cell migration. Original image: Human Protein Atlas (Image credit: Human Protein Atlas) [10,11]. Figure 3 is a local-visualization analysis of a representative field and is distinct from the five-case summary in Table 1.

In the pooled analysis of the five cases, 102 PD-1-positive and 465 PD-1-negative cells were identified. The median boundary distance was 52.6 px for PD-1-positive cells and 99.3 px for PD-1-negative cells, and PD-1-positive cells were significantly enriched near the cancer-nest boundary (one-sided Mann-Whitney U test, $p=1.93\times10^{-9}$). The mutual information between boundary distance R and PD-1 status G was $I(R;G) = 0.0302$ nats and was significant by stratified permutation testing in which PD-1 labels were shuffled within each image ($p$ = 0.00040). In addition, at the case level using case-wise median distances, the median R of PD-1-positive cells was on average 51.9 px shorter than that of PD-1-negative cells, which was significant by a one-sided Wilcoxon signed-rank test ($p$ = 0.0313; Figure 4).

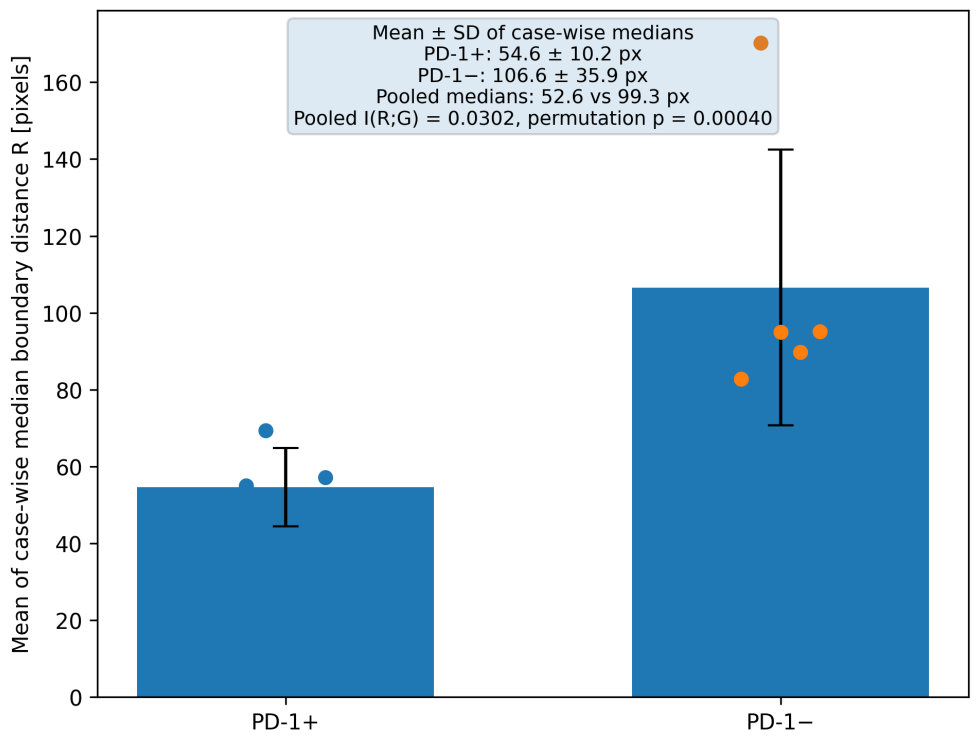

Figure 4. Mean case-wise median boundary distances for PD-1-positive and PD-1-negative cells across five cases. Dots indicate individual cases, and bars indicate mean ± SD. PD-1-positive cells showed shorter boundary distances than PD-1-negative cells (54.6 ± 10.2 px vs 106.6 ± 35.9 px). In the pooled analysis, the median boundary distances were 52.6 px and 99.3 px, respectively, and the mutual information between boundary distance R and PD-1 status G was $I(R;G)=0.0302$ nats, with $p=0.00040$ by stratified permutation testing.

**Table 1.** Boundary-distance information analysis in five HPA-derived PD-1 IHC cases. Distances are reported in pixels and information quantities in nats. Median $R$ +/− denotes the median boundary distance of PD-1-positive/negative cells. For the pooled analysis, labels were shuffled within each image using a stratified permutation test. Image source: Human Protein Atlas, PDCD1 in lung cancer (**Supplementary table 1**) [10,11]. Image credit: Human Protein Atlas. The patient/image ID, antibody ID, and direct image URL for each case should be provided in the supplementary table before preprint release.

| Case | PD-1+ | PD-1− | median R +/− | $I(R;G)$ | $p$ | $D_R$ | $D_{RG}$ |
|---|---|---|---|---|---|---|---|
| Case 1 | 30 | 42 | 55.0 / 82.8 | 0.1108 | 0.019 | 0.1679 | 0.2787 |
| Case 2 | 22 | 79 | 69.3 / 170.2 | 0.0975 | 0.0074 | 0.1007 | 0.1981 |
| Case 3 | 16 | 116 | 41.5 / 95.0 | 0.0404 | 0.191 | 0.1114 | 0.1518 |
| Case 4 | 16 | 155 | 50.2 / 89.7 | 0.0341 | 0.079 | 0.1849 | 0.2190 |
| Case 5 | 18 | 73 | 57.1 / 95.1 | 0.0476 | 0.274 | 0.1245 | 0.1720 |
| Pooled | 102 | 465 | 52.6 / 99.3 | 0.0302 | 0.00040 | 0.0965 | 0.1267 |

## 5. Discussion

### 5.1 Methodological Significance

This study compared the joint distribution of distance $R$ from a pathologically meaningful boundary and coarse-grained molecular expression state $G$ with $Q(R)P(G)$, where $Q(R)$ is derived from tissue geometry, and decomposed the total KLD into placement bias $D_R$ and state coupling $I(R;G)$. The decomposition itself follows the chain rule for KLD [1,2], whereas the distinguishing feature of the present framework is the introduction of a distance null distribution $Q(R)$ constructed from tissue geometry, thereby separating cellular localization relative to a pathological boundary from changes in expression-state composition.

If $D_R \approx 0$ and $I(R;G) > 0$, the overall cellular placement is close to the geometric null distribution while expression-state composition remains boundary-distance dependent. Conversely, if $D_R > 0$ and $I(R;G) \approx 0$, cells are positioned non-randomly relative to the boundary while state composition is not distance dependent. This distinction explicitly separates two effects that can otherwise be conflated by positive-cell density or median-distance comparisons.

### 5.2 Simulation Validation and Finite-Sample Effects

In the idealized simulation (Figure 1), $D_R$ became selectively positive for placement alone, whereas $I(R;G)$ became selectively positive for state coupling alone. The same separation behavior was retained in the semi-synthetic validation using fixed irregular tissue geometry reconstructed from an actual PD-1 image (Figure 2). The use of in silico tissue and spatial simulations with known ground truth to evaluate analytical behavior is established in spatial-omics method development, benchmarking, and power analysis [7-9]. Our simulations are not intended to reproduce biological reality; rather, they serve as mechanistic validation of the interpretation and statistical behavior of the proposed metrics.

Even under the Null condition, plug-in KLD and mutual information estimated from finite numbers of cells were slightly positive. Upward bias in finite-sample information estimation is well known [12,13], and it is therefore inappropriate to infer spatial organization by thresholding the absolute value of $I(R;G)$ alone. By permuting labels while holding cell positions and distances fixed, our procedure incorporates mutual information arising by chance under the same sample size and distance structure into the null distribution. This is particularly important in image analyses with limited numbers of cases and cells.

### 5.3 Comparison with Existing Spatial-Analysis Methods

A variety of metrics are used to study tumor microenvironments, including cell-cell distance, distance from tumor boundaries, Ripley-type statistics, and spatial entropy. FunSpace uses spatial entropy and mutual information across distance scales to quantify spatial heterogeneity in cellular composition [4]; Cottrazm analyzes the tumor microenvironment along a malignant-boundary-nonmalignant axis [5]; and Spatiopath evaluates cell-tumor spatial relationships under null hypotheses that account for tissue geometry [6]. In contrast, our framework explicitly combines the expression-state distribution $P(G \mid R)$, conditioned on a single pathologically defined boundary distance $R$, with the tissue-geometry null distribution $Q(R)$, and focuses on the additive decomposition of total divergence $D_{RG}$ into placement bias $D_R$ and state coupling $I(R;G)$.

Median distance provides an intuitive summary of the representative distance difference between positive and negative populations, but it does not fully describe cases in which composition changes across multiple distance ranges. By contrast, $I(R;G)$ averages the local KLD between $P(G \mid R = r)$ and the overall $P(G)$, weighted by $P(r)$, so both positive and negative states contribute. The distance-dependent positive fraction in Figure 3E and $J(R)$ in Figure 3A provide local visualizations of this change in composition.
The tumor-stroma boundary itself has been associated with immune-cell composition and immunotherapy response in spatial transcriptomic and multiplex imaging studies [14]. However, the present study did not analyze treatment-response data; therefore, the PD-1 boundary information quantity cannot yet be interpreted as a predictive biomarker for immune-checkpoint inhibitor response. Clinical utility should be treated as a hypothesis requiring independent-cohort validation.
The white arrows in Figure 3F visualize the distance gradient $\left|\frac{dJ}{dR}\right|$ of the local KLD projected toward the nearest cancer-nest boundary. Direction is geometrically defined and arrow length and thickness are rescaled for visualization. Thus, this display does not estimate a physical force field, a cell-migration vector, or causal information flow.
The present approach is therefore best regarded as complementary to, rather than a replacement for, existing point-process, autocorrelation, and distance-based analyses. Its specific role is to express changes in expression-state composition as an information quantity while treating the pathological boundary as an explicit conditioning variable.

**5.4 Limitations and Future Validation**
This study has several limitations. First, the empirical analysis is a small proof of concept based on five publicly available HPA IHC cases, and precision at the case level is limited. Because cases with larger cell counts may contribute disproportionately to pooled analyses, future studies should report case-level effect sizes and confidence intervals and use hierarchical models and independent cohorts for reproducibility assessment. Second, cancer-nest boundaries and PD-1-positive/negative cells were manually annotated; inter-observer reproducibility, automated or semi-automated segmentation, and sensitivity analysis to boundary perturbation are therefore needed. Third, distances were measured in pixels; calibration to micrometers is desirable when comparing images acquired under different conditions.
Fourth, the cancer-nest mask used in the semi-synthetic validation was reconstructed from manually drawn boundaries in the displayed image rather than from the original binary segmentation object. The final workflow should therefore preserve the original masks for each case and release them in a form suitable for reanalysis. Fifth, use of HPA images requires reporting the patient/image ID, antibody ID, atlas version, and direct URL for each case [11]. These identifiers should be verified before preprint submission as part of the reproducibility record.

**5.5 Relation to Information Thermodynamics and Conclusion**
In information thermodynamics, mutual information appears as an informational term that constrains work and entropy production when measurement and feedback are explicitly considered [15-18]. However, the present data consist of static IHC images and do not measure energy consumption, reaction rates, temporal directionality, or feedback operations. Accordingly, physical work or chemical potential cannot be inferred directly from $I(R;G)$ or from the local information gradient used here. Any connection to information thermodynamics should therefore remain a theoretical perspective for future studies incorporating time-resolved data or perturbation experiments.
In conclusion, a tissue-geometry-conditioned KLD decomposition distinguished cellular placement bias $D_R$ from boundary-distance-dependent expression-state coupling $I(R;G)$ in idealized and semi-synthetic simulations and in publicly available PD-1 IHC images. This framework provides a quantitative proof of concept for describing spatial organization of molecular expression states along pathological boundaries while separating it from placement effects.

## 6. Materials and Methods

### 6.1 Data Source, Image Selection, Annotation, and Boundary Distance

PDCD1 (PD-1) lung carcinoma IHC images were obtained from the publicly available Human Protein Atlas pages [10,11]. HPA states that its IHC pathology specimens are distinct from TCGA samples [11]. One representative field was selected from each of five cases, and cancer-nest boundaries were defined manually. PD-1-positive cells were annotated in red and negative cells in blue. For each cell, $x_i$ denotes the center coordinate and $\partial C$ the cancer-nest boundary; the shortest distance $R_i = d(x_i, \partial C)$ was calculated in pixels of the original image. For the local visualization in Figure 3, an inside-zero setting was used in which cells within cancer nests were treated as having reached the boundary and assigned $R = 0$. Because the aggregation settings differed from those used in the five-case analysis in Table 1, the cell counts in Figure 3 do not directly correspond to individual rows of Table 1.

$$R_i = d(x_i, \partial C)$$

In accordance with HPA usage and citation guidance, “Image credit: Human Protein Atlas” was included in Figure 2A and Figure 3F, which display original HPA images [11]. At preprint submission, the patient/image ID, antibody ID, HPA version, and direct image URL for each case should be provided in a supplementary table.

$$G_i = 1 \quad (PD - 1 IHCpositive), G_i = 0 \quad (PD - 1 IHCnegative)$$

### 6.2 Information Quantities and Geometry-Derived Null Distribution

Boundary distances $R_i$ were discretized into multiple distance bins, and the empirical joint distribution $P(r, g)$ was estimated. The mutual information between distance bin r and PD-1 state g was calculated as

$$I(R; G) = \sum_{r,g} P(r, g) \log \frac{P(r, g)}{P(r)P(g)}$$

Natural logarithms were used, and information was reported in nats. Because $G$ is binary, if $p_r = P(G = 1 \mid R = r)$ denotes the PD-1-positive fraction in distance bin r and $p = P(G = 1)$ the overall positive fraction, the same quantity can be written as

$$I(R; G) = \sum_{r} P(r) \left[ p_r \log \frac{p_r}{p} + (1 - p_r) \log \frac{1 - p_r}{1 - p} \right]$$

This expression is the distance-weighted average of the KLD between the local PD-1 state distribution $P(G \mid R = r)$ in each distance bin and the overall distribution $P(G)$.

Because cancer area and boundary length differed among images, the analyzable area $A_r$ in each distance bin was obtained from the image mask and the geometry-derived null distribution was defined as

$$Q(r) = \frac{A_r}{\sum_r A_r}$$

This allowed the placement bias $D_R$ of the cellular distance distribution to be separated from the distance dependence of PD-1 state $I(R; G)$.

$$D_R = D_{\mathrm{KL}}\big(P(R) \parallel Q(R)\big) = \sum_{r} P(r) \log \frac{P(r)}{Q(r)}$$

The total spatial-expression divergence was

$$D_{RG} = D_R + I(R; G)$$

and was calculated accordingly.

### 6.3 Statistical Analysis

The significance of $I(R;G)$ was evaluated using a permutation test in which cell positions and distances $R_i$ were held fixed while only PD-1 state labels $G_i$ were permuted. For case-level analyses, labels were permuted within each image. For the pooled analysis, a stratified permutation test was used in which labels were permuted only within each image to preserve image-specific cell counts and distance structures. The directional hypothesis that positive cells were closer to the boundary than negative cells was evaluated secondarily using a one-sided Mann-Whitney U test. Mutual information was calculated using natural logarithms and reported in nats. Because plug-in MI can be upwardly biased in finite samples [12,13], statistical inference was based on comparison with the permutation null rather than on the absolute observed MI value.

### 6.4 Image Processing, Local State-Information Gradient, and Numerical Computation

Image processing and numerical computation were performed in Python. Boundary masks were created from manually drawn boundaries, and cell centers were detected from connected components of the annotation colors. For the local visualization in Figure 3, the local positive fraction $q(R)$ was estimated from each cell's binary state $G$ and distance $R$ using a Gaussian kernel with bandwidth $h = 25$ pixels.

$$q(R) = \frac{\sum_i K_h\,(R_i - R)G_i}{\sum_i K_h\,(R_i - R)}$$

$$K_h(\Delta R) = \exp\left[-\frac{(\Delta R)^2}{2h^2}\right]$$

Local state information relative to the overall positive fraction $p = P(G = 1)$ was calculated as

$$J(R) = q(R)\log\frac{q(R)}{p} + [1 - q(R)]\log\frac{1 - q(R)}{1 - p}$$

$J(R)$ was numerically differentiated over the distance grid, and the magnitude of the local information gradient was defined as

$$F_{\text{info}}(R) = \left|\frac{dJ(R)}{dR}\right|$$

For positive cells outside cancer nests, arrows were drawn from the cell center toward the nearest cancer-nest boundary point, and arrow length, thickness, and head size were rescaled according to $F_{\text{info}}(R_i)$.

### 6.5 Idealized Simulation and Semi-Synthetic Validation

For the idealized toy simulation, a circular cancer nest with radius 0.25 was placed at the center of a unit square, and the region outside the circle was treated as the analyzable domain. Euclidean distances from analyzable pixels to the cancer boundary were discretized into 10 bins, and $Q(R)$ was constructed from the proportion of analyzable area in each bin. The placement distribution and binary state were generated as follows.

$$P_\alpha(r) = \frac{Q(r)e^{-\alpha r_{\text{scaled}}}}{\sum_{r'} Q\,(r')e^{-\alpha r'_{\text{scaled}}}}$$

$$P(G = 1 \mid R = r) = \text{logit}^{-1}[c + \beta(0.5 - r_{\text{scaled}})]$$

For each condition, c was adjusted so that the overall positive fraction matched the pooled empirical data (102/567). Four conditions were defined: Null ($\alpha = 0, \beta = 0$), Placement bias only ($\alpha = 2, \beta = 0$), State coupling only ($\alpha = 0, \beta = 3.5$), and Placement + state ($\alpha = 2, \beta = 3.5$). Finite-sample analyses evaluated $N = 50$, 100, 250, 567, 1000, and 2000 cells, including Monte Carlo analysis of Null bias and rejection rates for a permutation test in which $R$ was fixed and only $G$ labels were permuted. The parameter values shown in Figure 1 were selected to generate illustrative effect sizes and were not fitted to the empirical data.

To assess decomposition behavior in a geometry resembling real tissue, cancer-nest regions were reconstructed from manually drawn cancer-nest boundaries in an annotated PD-1 IHC image. Near-neutral dark boundary pixels were extracted from the displayed image, the major connected boundary components were retained, and large enclosed regions separated by the boundaries were treated as cancer-nest masks, with the central stromal region treated as analyzable. Because the original binary segmentation object was not contained in the manuscript file, this validation was based on geometry reconstructed from the displayed annotation. Euclidean distances from analyzable pixels to the cancer nests were calculated, and $Q(R)$ was constructed from the fraction of analyzable pixels in each distance bin. The reconstruction procedure is described in this section.

Semi-synthetic cellular placement and binary expression state were generated using the same equations as in the idealized simulation.

$$P_\alpha(r) = \frac{Q(r)e^{-\alpha r_{\text{scaled}}}}{\sum_{r'} Q\,(r')e^{-\alpha r'_{\text{scaled}}}}$$

$$P(G = 1 \mid R = r) = \text{logit}^{-1}[c + \beta(0.5 - r_{\text{scaled}})]$$

The parameter c was adjusted so that the overall positive fraction matched the pooled empirical data (102/567). Four conditions were evaluated: Null ($\alpha = 0$, $\beta = 0$), Placement bias only ($\alpha = 2$, $\beta = 0$), State coupling only ($\alpha = 0$, $\beta = 3.5$), and Placement + state ($\alpha = 2$, $\beta = 3.5$). Ground-truth $D_R$, $I(R;G)$, and $D_{RG}$ were calculated from the population distribution for each condition. In addition, $N = 567$ cells were sampled 2,000 times to evaluate the finite-sample behavior of the plug-in estimators. The α and β values were illustrative parameters chosen to visualize separation of effects and were not fitted to the empirical data.

### 6.6 Data Availability and Reproducibility

The HPA PDCD1 lung-cancer pages and images are publicly accessible resources [10,11]. The specific patient/image ID, antibody ID, and direct URL for each of the five cases used in this analysis should be added to the supplementary table before preprint submission. To ensure reproducibility, the analysis code should be released with the preprint as Code or in a public repository.

**Supplementary Table S1.** Human Protein Atlas image metadata for the five PDCD1 lung carcinoma cases(https://www.proteinatlas.org/ENSG00000188389-PDCD1/cancer/lung+cancer#img).
In accordance with the current HPA citation policy, the following items should be verified from the original images and completed before preprint submission. Identifiers that cannot be determined from the materials currently available in the manuscript have not been inferred.

| Case | HPA patient/image ID | Antibody ID | Atlas version |
|---|---|---|---|
| 1 | Patient id: 1907 | CAB076386 | 2026. June |
| 2 | Patient id: 4488 | CAB076386 | 2026. June |
| 3 | Patient id: 2222 | CAB076386 | 2026. June |
| 4 | Patient id: 2403 | CAB076386 | 2026. June |
| 5 | Patient id: 2003 | CAB076386 | 2026. June |

**Supplementary Note S1. Plug-In Estimators and Finite-Sample Bias**

A plug-in estimator is obtained by directly substituting empirical probabilities estimated from observed data into the definition of an information quantity when the true probability distribution is unknown.
If the true probability distribution were known, the mutual information between boundary distance *R* and expression state *G* used in this study would be defined as

$$I(R;G) = \sum_{r,g} P\,(r,g)\log\frac{P(r,g)}{P(r)P(g)}$$

In actual tissue images, however, the true $P(r,g)$, $P(r)$, and $P(g)$ are unknown. Let n(r,g) denote the observed number of cells in distance bin r and expression state g and N the total number of cells. The empirical joint distribution is estimated as

$$\hat{P}(r,g) = \frac{n(r,g)}{N}$$

and this empirical distribution is directly substituted into the mutual-information formula.

$$\hat{I}_{\text{plugin}}(R;G) = \sum_{r,g} \hat{P}\,(r,g)\log\frac{\hat{P}(r,g)}{\hat{P}(r)\hat{P}(g)}$$

Because the unknown P is replaced by its empirical estimate P̂ and 'plugged in' to the formula, this is called a plug-in estimator. The primary analysis in this study, in which the empirical joint distribution was constructed from observed boundary-distance bins and PD-1 expression states and then used to calculate $I(R;G)$, is a plug-in-type estimation.

Positive Finite-Sample Bias
With plug-in mutual information, the estimate tends to take a small positive value when only a finite number of observations are available, even if the true mutual information is zero. Systematic errors in information estimates caused by limited sampling have long been recognized. Panzeri and Treves analyzed systematic error arising from limited sampling, and Panzeri et al. further described the upward bias of plug-in information estimators based directly on empirical frequencies [12,13].
For example, suppose that boundary distance $R$ and expression state $G$ are independent in the true population, so that

$$R \perp G, \qquad I(R;G) = 0$$

Even under this condition, random sampling of a finite number of cells causes the positive fraction to fluctuate across distance bins. Even if the true positive fraction is 18% at every distance, a finite sample may yield values such as 15%, 21%, 17%, 20%, and so on. Because these chance fluctuations are directly incorporated into the empirical distribution used in the mutual-information formula,

$$\hat{I}_{\text{plugin}}(R;G) > 0$$

may occur. Therefore, observing a small positive $I(R;G)$ does not by itself demonstrate true spatial dependence.

Confirmation in the Present Simulations

In the toy simulation, although the population ground truth was set to $I(R;G) = 0$ under the Null condition, the mean plug-in estimate obtained from finite samples of $N = 567$ cells was approximately 0.008 nats. As the number of cells increased, this positive bias gradually approached zero. The upper panel of Figure 1D illustrates this finite-sample effect.
These results indicate that the pooled value $I(R;G) = 0.0302$ nats observed in this study should not be interpreted as significant on the basis of its absolute magnitude alone. We therefore used a permutation test in which cell positions and boundary distances were held fixed while only PD-1-positive/negative labels were randomized. Under this procedure, mutual information arising by chance from finite sampling under the same cell number and distance structure is included in the null distribution, allowing assessment of whether the observed $I(R;G)$ exceeds sampling variation.

Interpretive Considerations

Interpretation of $I(R;G)$ in this study therefore requires distinguishing among (i) the plug-in estimate itself, obtained by direct substitution of the empirical distribution; (ii) positive finite-sample bias; and (iii) the permutation null generated by label randomization. Statistical significance is based not on the mere positivity of the observed value but on the degree to which it departs from the null distribution under the same sample size and tissue geometry.